\documentclass[11pt]{article}
\begin{document}
\newcommand{\hh}{$\mbox{H}^2$}
\newcommand{\up}{``$\uparrow $''}

\vspace*{1.7in}
\begin{center}\large{\bf ``Quantum-states-as-information'' meets Wigner's 
       friend:\\ A comment on Hagar and Hemmo}\\
\vspace{1cm}
\normalsize\ J. Finkelstein\footnote[1]{
        Participating Guest, Lawrence Berkeley National Laboratory\\
        \hspace*{\parindent}\hspace*{.5em}
        Electronic address: JLFINKELSTEIN@lbl.gov}\\
        Department of Physics\\
        San Jos\'{e} State University\\San Jos\'{e}, CA 95192, U.S.A
\end{center}
\begin{abstract}
This is a comment on the paper by Hagar and Hemmo (quant-ph/0512095)
in which they suggest that information-theoretic approaches to quantum
theory are incomplete.
\end{abstract}
\newpage
Quantum theory is a remarkably successful theory.  There is little
doubt
as to what quantum theory {\em says}, at least in simple situations;
predictions of quantum theory have been experimentally verified with
great accuracy, and no prediction of quantum theory is known to be incorrect.
On the other hand, there is no consensus as to what quantum theory
{\em means}.

In a recent article \cite{hh}, Hagar and Hemmo (henceforth \hh) have
discussed the position, especially as set forth by Bub \cite{Bu} and
by  Fuchs \cite{CF}, that quantum theory should be considered to be
about quantum information, rather than about quantum systems.
 \hh\ suggest that the story of
``Wigner's friend'' \cite{W} would present a difficulty for this
position, and in this note I wish to comment on this suggestion. To
avert a possible misunderstanding, I should point out that \hh\ also
emphasize that the experimental successes of quantum theory do not
rule out other theories which are, at least
in principle, empirically distinguishable from quantum theory
(they cite the model of Ghirardi, Rimini, and Weber \cite{GRW} as an
example), 
but that is not the aspect of their article upon which I will comment.
For the purpose of this note, I will  take quantum theory to be
empirically completely correct.

Here is a version of the story of Wigner's friend, similar to the
version told by \hh : A friend of Wigner's enters a room, which is then
closed; Wigner remains outside.  Inside the room is an electron, which
has been prepared with its spin along the $x$-axis, so that
\begin{equation}|\mbox{initial}\rangle _{e} = \frac{1}{\sqrt{2}}[|\uparrow
    \rangle _{e} +|\downarrow
    \rangle _{e}]. \label{est}
\end{equation}
Here subscript $e$ refers to the electron, and $|\uparrow
\rangle _{e}$ ($|\downarrow\rangle _{e}$) represent states in
which its spin is parallel (anti-parallel) to the $z$-axis.
The room also contains a machine, whose operation I will describe 
below, but which is
initially switched off and thus completely inert.  After the room has
been closed, Friend measures the component of spin of the electron along
the $z$-axis.  Wigner has been told that Friend would measure
that spin component, but he does not know the result of the
measurement, so Wigner assigns the following state-vector to the
contents of the room:
\begin{equation} |\Psi \rangle = \frac{1}{\sqrt{2}}[|\uparrow
    \rangle _{e}|\mbox{sees} \uparrow\rangle _{R} +|\downarrow
    \rangle _{e}|\mbox{sees} \downarrow\rangle _{R}], \label{st1}
\end{equation}
where system $R$ consists of the part of Friend which remembers the result
of the measurement, together with anything else (other parts of
Friend, the measuring apparatus, the air in the room...) with which
that part may have become entangled. 

I will discuss the two cases in which Friend does, or does not, switch on
the machine after he has measured the spin.  If he does not, then the machine
remains inert, and so has no effect on the rest of our story.  
If the machine is switched on, here is what happens: there is a
pointer mounted on the outside of the machine, which can point either to a
mark ``Y'' or to a mark ``N''; when switched on, the machine
interacts with the combined system $e\otimes R$ with the result that 
if that system
is described by the state-vector given in eq.\ \ref{st1} (as in fact
it is) the pointer points
to ``Y'', but if the state-vector were orthogonal to that given in 
eq.\ \ref{st1}, it would
point to ``N''.  Of course such a  machine is completely impossible in
practice \cite{whyR}, but having made the assumption that quantum
theory gives correct answers in {\em all} circumstances, we are
entitled to consider what would happen if such a machine were actually
to exist.  And what would happen is that, if Wigner enters the room
after the machine has been switched on, he will surely see that the
pointer indicates 
``Y''.  In contrast to this, in a ``collapse theory'' such as that of 
\cite{GRW}, after the spin is measured by Friend, the state-vector of
$e\otimes R$ quickly collapses to either the first or the second term
on the right-hand side of eq.\ \ref{st1}, which would imply that the
pointer would indicate ``Y'' only with probability 
\footnotesize $\frac{1}{2}$. \normalsize 
Thus the machine, if it were to exist, could distinguish between a
collapse and a no-collapse theory. 

Let us now consider what Friend thinks of all this.  Suppose that
Friend determines the $z$-component of spin of the electron by looking at
a measuring device which displays either a \up\ or a 
``$\downarrow $'', and  that in fact it is the
\up\ which he sees displayed.  He now wants to predict how the
machine's pointer will point when he switches it on.  He can certainly 
calculate (just as I have done above) that Wigner will assign the
state-vector written in eq.\ \ref{st1}, and that means that he knows that
Wigner predicts with certainty that the pointer will indicate ``Y'';
under the assumption that results predicted by quantum theory are
correct, this means that the pointer {\em will} indicate ``Y'' \cite{look}. 
In fact, {\em even before Friend entered the room,} both he and Wigner knew
that, as long as Friend would indeed measure the electron's spin and
then turn on the machine, the pointer would indicate ``Y''.

So far, so good.  However, \hh\ suggest that information-theoretic
approaches to quantum theory put Friend in danger of making
``inconsistent predictions''.
  Their argument leading to the (apparent) inconsistency is
that Friend, having seen the result ``$\uparrow $'', should
``collapse'' the state-vector given in eq.\ \ref{st1} to get 
\begin{equation} \Psi_{\mbox{\footnotesize collapsed\normalsize}} 
= |\uparrow
    \rangle _{e}|\mbox{sees} \uparrow\rangle _{R}, \label{st2}
\end{equation}
and then should use this collapsed state-vector to predict that the
pointer would indicate ``Y'' only with probability 
\footnotesize $\frac{1}{2}$; \normalsize
this would indeed be inconsistent with the prediction that the pointer
would indicate ``Y'' with probability 1.  \hh\ argue further that
the
only way to avoid this inconsistency while maintaining an
information-theoretic approach to quantum theory
is to make assumptions about what
would happen to Friend's memory during its interaction with the
machine; since these assumptions lie outside of that approach,
\hh\ conclude that an information-theoretic approach, if it is to
avoid inconsistency, must be incomplete.
\hh\ write that, in the information-theoretic approach, ``the
assignment of quantum states becomes (in some circumstances)
ambiguous,'' i.e.\ that Friend might assign to the $e\otimes R$ system
either of the state-vectors $\Psi $ (written here in eq.\ \ref{st1}) or 
$\Psi_{\mbox{\footnotesize collapsed\normalsize}}$ (eq.\ \ref{st2}). I
will argue below that {\em neither} of these assignments would be correct.
It is true that if
Friend, before he switched on the machine, were to inform Wigner that
he had observed the result \up , Wigner would base his
prediction for the pointer on the state-vector 
$\Psi_{\mbox{\footnotesize collapsed\normalsize}}$ (and
in that case this new prediction would be the correct one, since the
machine does not interact with Wigner).  But given that Friend does
not so inform Wigner, the state-vector $
\Psi_{\mbox{\footnotesize collapsed\normalsize}}$  would not be
assigned by either one of them, so it is not clear that there  is
any potential inconsistency to be avoided.

In any case, there is a way to exhibit some unusual aspects of
Friend's situation, without having to consider at all what might happen
to him or to his memory during the interaction with the machine.  
Suppose that Friend
does {\em not} turn on the machine, but instead makes a second
measurement of the $z$-component of the electron's spin.  If the machine
were not present, it is an elementary result of quantum theory
that this second measurement would have the same result as the first 
(assuming of course that there were no stray magnetic fields which
might cause the electron to precess).
Surely the mere presence of the machine, if it is never switched on, 
would not alter this result.  Therefore Friend knows that this
second measurement of spin would surely have the result ``$\uparrow $''

Thus we see that, at a time after he has made the initial measurement
of the $z$-component of spin, but when he has not yet switched on the
machine
\begin{description}
\item{A)}  Friend knows that if he switches on the machine, the 
result will be ``Y''.
\item{B)} Friend knows that if he does  {\em not} switch on the machine, 
but instead makes a second measurement of the spin, the result will be \up .
\end{description}
Because the operator corresponding to the observable measured by the
machine does not commute with the spin operator for the electron,
these two statements do not imply that Friend knows the result he
would see if he first switched on the machine, and then made a
second measurement of spin.  Nevertheless he is predicting with
certainty the results of two incompatible measurements. Note that this 
conclusion does not depend on any identification of state-vectors as
representing information, nor indeed on any commitment to what quantum
theory {\em means}; statements A) and B) are just what quantum theory
{\em says}.  Statement B) is an elementary-textbook statement about a
repeated measurement on a microscopic system, while statement A)
requires us to trust what quantum theory says about
(impossible-in-practice) measurements on macroscopic systems. 

Statements A) and B) together imply  that Friend cannot 
ascribe {\em any} quantum state to the system $e\otimes R$; 
there simply is no quantum state, neither pure nor mixed, which is
compatible with statements A) and B) \cite{proof}  
The information which Friend has about the results of future measurements
on the $e\otimes R$ system does not correspond to any quantum state of
that system.

\vspace{1cm}
Acknowledgement: I would  like to acknowledge the hospitality of the
Lawrence Berkeley National Laboratory, where this work was done.


\begin{thebibliography}{9}
\bibitem{whyR} It would also be impossible in principle, unless
         the machine were allowed to interact with anything with which
         Friend might be entangled, which is why I chose to define the
         system $R$ in the way I did.  A more-usual treatment would be
         to not mention this issue under the (almost-surely correct)
         assumption that the reader already understands it.
\bibitem{look} We could even go further and state that Friend
           therefore knows that if he himself should look at the
           pointer after turning on the machine, he would see it
           indicating ``Y''.
\bibitem{proof} For a simple proof of this, let $P_{\Psi}$ be the
  projector $|\Psi \rangle \langle \Psi |$ (where $|\Psi \rangle $ is
  given in eq.\ \ref{st1}), let $P_{\uparrow }$ be the projector
  $(|\uparrow \rangle _{e}\langle \uparrow|) \otimes \mbox{I}_{R}$, and
  suppose there were a density operator $\rho$ for the $e\otimes R$
  system.  Then statement A) would require
\begin{equation}
\mbox{Probability(``Y'')} = \mbox{Tr}(P_{\Psi}\rho) = 1, \label{e1}
\end{equation}
while statement B) would require
\begin{equation}
\mbox{Probability(``$\uparrow$'')} = \mbox{Tr}(P_{\uparrow}\rho) = 1. 
\label{e2}
\end{equation}
Eq.\ \ref{e1} has the unique solution $\rho =|\Psi \rangle \langle
\Psi |$ (again, $|\Psi \rangle $ from eq.\ \ref{st1}), but with that
solution for $\rho $, eq.\ \ref{e2} reads $\frac{1}{2} = 1.$
\end{thebibliography}
\end{document}